\documentstyle[amsfonts,pra,aps]{revtex}
\newcommand{\be}{\begin{equation}}
\newcommand{\ee}{\end{equation}}
\newcommand{\br}{\begin{eqnarray}}
\newcommand{\er}{\end{eqnarray}}
\newcommand{\bd}{\begin{displaymath}}
\newcommand{\ed}{\end{displaymath}}
\newcommand{\nn}{\nonumber}
\newcommand{\bib}{\bibitem}
\newcommand{\bfig}{\begin{figure}}
\newcommand{\efig}{\end{figure}}
\def\alf{\alpha}
\def\th{\theta}

\def\lam{\lambda}
\def\alf{\alpha}
\def\rpar{\right)}
\def\lpar{\left(}
\def\rbk{\right]}
\def\lbk{\left[}
\def\rbr{\right\}}
\def\lbr{\left\{}
\def\lb{\label}
\def\im{\mbox{$\dot{\imath}$}}
\def\tr{\mbox{${\rm Tr}$}}
\def\ro{\mbox{\boldmath $\rho$}}
\def\lbf{{\bf \Lambda}}
\def\gb{{\bf \Gamma}}
\def\rg{\rangle}
\def\lg{\langle}

\def\wtil{\widetilde}
\def\rf#1{(\ref{#1})}
\begin{document}
\draft
\title{Marginal and correlation distribution functions in the squeezed-states 
representation}
\vspace{1.0in}
\author{Marcelo A. Marchiolli\thanks{march@if.sc.usp.br}}
\address{Instituto de F\'{\i}sica de S\~ao Carlos, Universidade de S\~ao 
         Paulo, \\
         Caixa Postal 369, 13560-970 S\~ao Carlos, SP, Brazil}
\author{Salomon S. Mizrahi\thanks{salomon@power.ufscar.br}  
        and Victor V. Dodonov\thanks{vdodonov@power.ufscar.br}} 
\address{Departamento de F\'{\i}sica, Universidade Federal de S\~ao Carlos, \\
         Rod. Washington Luiz km 235, 13565-905 S\~ao Carlos, SP, Brazil}
\date{\today}
\maketitle
\begin{abstract}
Here we consider the Husimi function $P$ for the squeezed states and calculate 
the marginal and correlation distribution functions when $P$ is projected onto 
the photon number states. According to the value of the squeezing parameter 
one verifies the occurence of oscillations and beats as already appointed in 
the literature. We verify that these phenomena are entirely contained in the 
correlation function. In particular, we show that since the Husimi and its 
marginal distribution functions satisfy partial differential equations where 
the squeeze parameter plays the role of time, the solutions (the squeezed 
functions obtained from ``initial'' unsqueezed functions) can be expressed by 
means of kernels responsible for the ``propagation'' of squeezing. From the 
calculational point of view, this method presents advantages for calculating 
the marginal distribution functions (compared to a direct integration over one 
of the two phase-space variables of $P$) since one can use the symmetry 
properties of the differential equations.
\end{abstract}
\vspace{5mm}
%
\section{Introduction}

In the end of the eighties, the oscillations of the photon distribution 
function (DF) of high energy squeezed and correlated states were discovered in 
\cite{s1,s2}; the authors of \cite{s1} studied the oscillatory behavior of 
the single-mode squeezed-state Husimi function (HF) projected into photon 
number states, $P_{n}(p,q;\lam,\phi) = | \lg n | p,q; \lam, \phi \rg |^{2}$, 
where $p$ and $q$ are the space variables associated with the two quadratures 
of a monochromatic electromagnetic (EM) field, $\lam$ is the squeezing 
parameter and $\phi$ is a rotation angle in phase space. They suggested that 
for $p=0$, $q = 7 \sqrt{2}$, $\phi=0$ and a fixed $\lam = 21$, such 
oscillatory behavior can be explained in terms of quantum interference effects 
and were taken as a signature of a nonclassical state. More recently, Gagen 
\cite{s3} generalized this study to incorporate interference structures in the 
Bohr-Sommerfeld trajectories associated with a superposition of quantum states. 
On the other hand, Dutta {\em et al}. \cite{s4} verified an additional feature present 
in $P_{n}(p,q;\lam,\phi)$, when plotted as function of $n$: this distribution 
exhibits the structure of beats (collapses and revivals) at large values of 
$n \; (\geq 10)$ for $\lam = 201$, $p^{2} + q^{2} = 98$ and $\phi \approx \pi
/2$. Compared to the value $\lam$ used in \cite{s1}, the high value for $\lam$ 
is crucial for the occurence of beats. These oscillations were attributed to 
interference effects in phase space \cite{s4}; however, since a detailed 
explanation about the beats has not been presented until the present time, we 
judge that they deserve a deeper investigation. 

Recently, Chountasis and Vourdas \cite{s5} showed that the Weyl function is 
an important tool for quantum interference effects. In particular, they 
studied the Wigner and Weyl functions for a superposition of $m$ quantum 
states $| s_{i} \rg$, where each function is decomposed into diagonal and 
nondiagonal terms, where the nondiagonal term is responsible for the 
interference between the states $| s_{i} \rg$. Adopting a different approach 
and using the formalism developed in \cite{s6}, and in order to shed more 
light on the origin of the beats predicted in \cite{s4}, here we decompose the
squeezed state HF into three functions: the two marginal (one for $p$ and the 
other for $q$) and the correlation distribution functions (MDF and CDF, 
respectively). Our results corroborate the phase-space interference concept 
and complement the graphical treatment proposed by Mandal \cite{s7}.

This paper is organized as follows. In section II we discuss the solution of 
the differential pseudo-diffusion equation (see reference \cite{s6}), and in 
section III we define the phase space MDF and CDF. The formal and numerical 
results are given in section IV, where we show that oscillations and beats are present 
in the CDF function. Section V is dedicated to a summary and conclusions. Two 
appendices are also presented, containing calculational details. Appendix A 
contains the main steps to calculate the MDFs by direct integration, and in 
appendix B we calculate the CDF.

\section{The pseudo-diffusion equation and its solutions}

The mapping of the statistical operator $\ro$ (describing a state of the EM 
field) on the squeezed-states representation (SSR) permits us to write the HF 
$P$ as follows \cite{s6}:
\be
\lb{e1}
P(p,q;\lam,\phi) = \tr \lbk \ro {\bf \Pi}(p,q;\zeta) \rbk = \tr \lbk \ro_{r} 
{\bf \Pi}(p_{r},q_{r};\lam) \rbk = P_{r}(p_{r},q_{r};\lam) \; , 
\ee
where
\be
\lb{e2}
{\bf \Pi}(p,q;\zeta) = |pq; \zeta \rg \lg pq; \zeta | = {\bf R}(\phi/2) 
{\bf \Pi}(p_{r},q_{r};y) {\bf R}^{\dagger}(\phi/2) 
\ee
is a projection operator, ${\bf R}(\phi/2) = \exp \lpar \frac{\im \phi}{2} \;
{\bf a}^{\dagger} {\bf a} \rpar$ is the rotation operator, $q_{r} = q \cos 
(\phi/2) + p \sin (\phi/2)$ and $p_{r} = p \cos (\phi/2) - q \sin (\phi/2)$ 
are the rotated phase space variables expressed in terms of the old ones, 
$\ro_{r} = {\bf R}^{\dagger}(\phi/2) \; \ro \; {\bf R}(\phi/2)$ and $\lam 
\equiv e^{-2y} \; (0 < \lam < \infty)$. Now, if one considers the mixed state 
$\ro = \sum_{n=0}^{\infty} p_{n} |n \rg \lg n|$, diagonal in the Fock basis, 
$\ro_{r}$ will be invariant under rotations since ${\bf R}(\phi/2) | n \rg = 
e^{\im n \phi/2} |n \rg$. Consequently, the associated HF is given by $P(p,q;
\lam,\phi) = P(p_{r},q_{r};\lam)$. This relation is useful in the sense that 
it is sufficient to consider the calculation of the unrotated distribution 
$P(p,q;\lam)$, with variables changed from $(p,q)$ to $(p_{r},q_{r})$ in the 
final result, respectively. The number state $|n \rg \lg n |$ is a typical 
example where this relation can be used. In \cite{s6} we demonstrated that 
$P(p,q;\lam,\phi)$ satisfies the partial differential equation
\be
\lb{e3}
\gb(p,q;\lam,\phi) P(p,q;\lam,\phi) = 0 \; , 
\ee
where
\br
\lb{e4} 
\gb(p,q;\lam,\phi) &=& \frac{\partial}{\partial \lam} - \frac{1}{4 \lam^{2}} 
\lbr \lbk \lam^{2} \cos^{2} (\phi/2) - \sin^{2} (\phi/2) \rbk 
\frac{\partial^{2}}{\partial p^{2}} + \lbk \lam^{2} \sin^{2} (\phi/2) - 
\cos^{2} (\phi/2) \rbk \frac{\partial^{2}}{\partial q^{2}} \right. \nn \\
& & - \left. (\lam^{2} + 1) \sin \phi \; \frac{\partial^{2}}{\partial q 
\partial p} \rbr 
\er
is a linear differential operator. For $\phi=0$, equation (\ref{e3}) is
similar to the diffusion equation in two dimensions where the parameter $\lam$ 
plays the role of time. In this situation, since the diffusion coefficients 
have opposite signs, the equation describes a diffusive (infusive) process in 
the $p$ ($q$) variable. For this reason, it has been called the 
{\em pseudo-diffusion equation} \cite{s8,s9}.

Here we consider the formal solution of equation (\ref{e3}), obtained by the 
Fourier transform (FT) method, written as an integral equation with the kernel 
depending on the squeeze and rotation parameters, 
\br
\lb{e5}
P(p,q;\lam,\phi) &=& \int_{- \infty}^{\infty} \frac{d\xi d\eta}{2 \pi} \; 
e^{\im (\eta p - \xi q)} K(\xi,\eta;\lam,\phi) \widetilde{P}(\xi,\eta) \nn \\
&=& \int_{- \infty}^{\infty} \frac{d\xi d\eta}{2 \pi} \; e^{\im (\eta p_{r} - 
\xi q_{r})} K(\xi,\eta;\lam,0) \widetilde{P}_{r}(\xi,\eta) \nn \\
&=& P_{r}(p_{r},q_{r};\lam) \; .
\er
The kernel is
\be
\lb{e6}
K(\xi,\eta;\lam,\phi) = \exp \lpar - \frac{\lam - 1}{4 \lam} \lbr [ \lam 
\sin^{2} (\phi/2) - \cos^{2} (\phi/2)] \xi^{2} + [ \lam \cos^{2} (\phi/2) - 
\sin^{2} (\phi/2)] \eta^{2} + (\lam + 1) \sin \phi \; \xi \eta \rbr \rpar 
\ee
and $\wtil{P}(\xi,\eta)$ is the FT of the HF $P(p,q)$ for $\lam =1$ and $\phi 
=0$. We notice that $K(\xi,\eta;\lam,\phi)$ is an {\em unbounded} function,  
responsible for the squeezing `propagation' of an `initial' function $P(p,q)$ 
to $P(p,q;\lam,\phi)$ for any values of $\lam$ and $\phi$ in their domain. 
Thus the existence of a `propagated' $P(p,q;\lam,\phi)$ depends on the 
functional form of $\wtil{P}(\xi,\eta)$, since the integral in the first line 
in (\ref{e5}) for $\wtil{P}(\xi,\eta) =$ constant does not exist. From the 
pseudo-diffusion equation \rf{e3} and the linear differential operator $\gb$,
equation (\ref{e5}) shows the following symmetry properties
\be
\lb{e7}
P(p,q;\lam,\phi) = P(q,-p;\lam,\phi \pm \pi) = P(q,p;\lam^{-1},-\phi)
\ee
and
\be
\lb{e8}
K(\xi,\eta;\lam,\phi) = K(\eta,-\xi;\lam,\phi \pm \pi) = K(\eta,\xi;\lam^{-1},
-\phi) \; . 
\ee

Now, our aim is to show that the Glauber-Sudarshan distribution $P^{c}(p,q;
\lam,\phi)$ and HF $P(p,q;\lam,\phi)$ are related by
\be
\lb{e9}
P^{c}(p,q;\lam,\phi) = \lbf(p,q;\lam,\phi) P(p,q;\lam,\phi) = P(p,q;-\lam,
\phi) \; , 
\ee
with
\be
\lb{e10}
\lbf(p,q;\lam,\phi) \equiv \exp \lbk - \frac{1}{2} \lpar \lam 
\frac{\partial^{2}}{\partial p_{r}^{2}} + \lam^{-1} \frac{\partial^{2}}
{\partial q_{r}^{2}} \rpar \rbk \; , 
\ee
which can also be written as
\bd
\lbf(p,q;\lam,\phi) = \exp \lbr - \frac{1}{2 \lam} \lbk [ \lam^{2} \cos^{2} 
(\phi/2) + \sin^{2} (\phi/2) ] \frac{\partial^{2}}{\partial p^{2}} + [ 
\lam^{2} \sin^{2} (\phi/2) + \cos^{2} (\phi/2) ] \frac{\partial^{2}}{\partial 
q^{2}} - (\lam^{2} - 1) \sin \phi \; \frac{\partial^{2}}{\partial p \partial 
q} \rbk \rbr \; ,
\ed
since
\br
\frac{\partial^{2}}{\partial p_{r}^{2}} &=& \sin^{2} (\phi/2) 
\frac{\partial^{2}}{\partial q^{2}} + \cos^{2} (\phi/2) \frac{\partial^{2}}
{\partial p^{2}} - \sin \phi \frac{\partial^{2}}{\partial q \partial p} \; , 
\nn \\ 
\frac{\partial^{2}}{\partial q_{r}^{2}} &=& \cos^{2} (\phi/2) 
\frac{\partial^{2}}{\partial q^{2}} + \sin^{2} (\phi/2) \frac{\partial^{2}}
{\partial p^{2}} + \sin \phi \frac{\partial^{2}}{\partial q \partial p} \; . 
\nn
\er
Applying the differential operator $\lbf$ on (\ref{e5}), we get 
\br
\lb{e11}
\lbf(p,q;\lam,\phi) P(p,q;\lam,\phi) &=& \int_{- \infty}^{\infty} \frac{d\xi 
d\eta}{2\pi} \lbk \lbf(p,q;\lam,\phi) \; e^{\im (\eta p_{r} - \xi q_{r})} \rbk 
K(\xi,\eta;\lam,0) \widetilde{P}_{r}(\xi,\eta) \nn \\
&=& \int_{- \infty}^{\infty} \frac{d\xi d\eta}{2\pi} \; e^{\im (\eta p_{r} - 
\xi q_{r})} \underbrace{e^{\frac{1}{2} (\lam^{-1} \xi^{2} + \lam \eta^{2})} 
K(\xi,\eta;\lam,0)}_{K(\xi,\eta;-\lam,0)} \widetilde{P}_{r}(\xi,\eta) \nn \\
&=& \int_{- \infty}^{\infty} \frac{d\xi d\eta}{2\pi} \; e^{\im (\eta p_{r} - 
\eta q_{r})} K(\xi,\eta;-\lam,0) \widetilde{P}_{r}(\xi,\eta) \nn \\
&=& P(p,q;-\lam,\phi) \; . 
\er
The second equality is obtained using the following relation
\br
\lbf(p,q;\lam,\phi) \; e^{\im (\eta p_{r} - \xi q_{r})} &=& 
\sum_{k=0}^{\infty} \frac{(-1)^{k}}{2^{k} k!} \lpar \lam^{-1} 
\frac{\partial^{2}}{\partial q_{r}^{2}} + \lam \frac{\partial^{2}}{\partial 
p_{r}^{2}} \rpar^{k} e^{\im ( \eta p_{r} - \xi q_{r})} \nn \\
&=& \sum_{k=0}^{\infty} \frac{1}{k!} \lbk \frac{1}{2} \lpar \lam^{-1} \xi^{2} 
+ \lam \eta^{2} \rpar \rbk^{k} e^{\im (\eta p_{r} - \xi q_{r})} \nn \\
&=& e^{\frac{1}{2} (\lam^{-1} \xi^{2} + \lam \eta^{2})} \; e^{\im (\eta p_{r} 
- \xi q_{r})} \nn
\er
and, from the definition of the kernel $K(\xi,\eta;\lam,0)$, we conclude that
\bd
e^{\frac{1}{2} (\lam^{-1} \xi^{2} + \lam \eta^{2})} K(\xi,\eta;\lam,0) = 
e^{\frac{1}{4} (\lam + 1) \lpar \eta^{2} + \lam^{-1} \xi^{2} \rpar} = 
K(\xi,\eta;-\lam,0) \; . 
\ed
Thus, the DF $P^{c}(p,q;\lam,\phi)$ is directly obtained by changing the 
signal of the squeezing parameter $\lam \to -\lam $ in the HF. Consequently, 
this result shows that $P^{c}_{n}(p,q;\lam,\phi)$ does not exist as a bounded 
function for the number states, however it exists as an ultradistribution.  
Equation (\ref{e9}) is a generalization of previous results obtained in 
\cite{s9,s10} for $\phi = 0$.

\section{Marginal and correlation distribution functions}

The HF $P(p,q;\lam,\phi)$ can be written as a sum of two terms: the first is 
the product of the two MDF, in $q$ and $p$, and describes the noncorrelated 
part; the second term is the CDF and contains the phase-space correlations 
\cite{s6}. So, the HF (\ref{e1}) can be written as 
\be
\lb{e12}
P(p,q;\lam,\phi) = R(p;\lam,\phi) Q(q;\lam,\phi) + C(p,q;\lam,\phi) \; ,
\ee
where
\be
\lb{e13}
Q(q;\lam,\phi) = \int_{- \infty}^{\infty} \frac{dp}{\sqrt{2 \pi}} \;
P(p,q; \lam,\phi) \; ,
\ee
and
\be
\lb{e14}
R(p;\lam,\phi) = \int_{- \infty}^{\infty} \frac{dq}{\sqrt{2 \pi}} \;
P(p,q; \lam,\phi)  
\ee
are the MDFs, and $C(p,q;\lam,\phi)$ is the CDF. However, the calculation of 
(\ref{e13}) and (\ref{e14}) by direct integration displays difficulties when 
$\phi \neq 0$ (see appendix A). Thus, the aim of this section is to show that 
expressions for the MDFs can be obtained in a much simpler way by using the 
formalism of section II.

Substituting the right hand side (RHS) of the first line of equation 
(\ref{e5}) into the integrands of equations (\ref{e13}) and (\ref{e14}), and 
then carrying out the integrations we get
\be
\lb{e15}
Q(q;\lam,\phi) = \int_{- \infty}^{\infty} \frac{d \xi}{\sqrt{2 \pi}} \;
e^{- \im \xi q} k_{Q}(\xi;\lam,\phi) \widetilde{Q} (\xi) 
\ee
and
\be
\lb{e16}
R(p;\lam,\phi) = \int_{- \infty}^{\infty} \frac{d \eta}{\sqrt{2 \pi}} \;
e^{\im \eta p} k_{R}(\eta;\lam,\phi) \widetilde{R} (\eta) \; , 
\ee
where
\be
\lb{e17}
k_{Q}(\xi;\lam,\phi) = \exp \lbr - \frac{\lam - 1}{4 \lam} \lbk \lam 
\sin^{2} (\phi/2) - \cos^{2} (\phi/2) \rbk \xi^{2} \rbr 
\ee
and
\be
\lb{e18}
k_{R}(\eta;\lam,\phi) = \exp \lbr - \frac{\lam - 1}{4 \lam} \lbk \lam 
\cos^{2} (\phi/2) - \sin^{2} (\phi/2) \rbk \eta^{2} \rbr 
\ee
are the reduced kernels responsible for the `propagation' of the squeezing.
The functions $\widetilde{Q}(\xi)$ and $\widetilde{R}(\eta)$ are the
respective FT of the Husimi functions $Q(q)$ and $R(p)$ for $\lam = 1$ 
(absence of squeezing). Futhermore, the equations (\ref{e15}) and (\ref{e16}) 
are solutions of the partial differential equations
\br
\lb{e19}
\lbk \frac{\partial}{\partial \lam} - \frac{\lam^{2} \sin^{2} (\phi/2) -
\cos^{2} (\phi/2)}{4 \lam^{2}} \frac{\partial^{2}}{\partial q^{2}} \rbk
Q(q;\lam,\phi) &=& 0 \; , \\
\lb{e20}
\lbk \frac{\partial}{\partial \lam} - \frac{\lam^{2} \cos^{2} (\phi/2) -
\sin^{2} (\phi/2)}{4 \lam^{2}} \frac{\partial^{2}}{\partial p^{2}} \rbk
R(p;\lam,\phi) &=& 0 \; .
\er
In analogy to (\ref{e8}), the reduced kernels $k_{Q}$ and $k_{R}$ have the
symmetry properties
\be
\lb{e21}
k_{Q(R)}(x;\lam,\phi) = k_{R(Q)}(x;\lam^{-1},\phi) = k_{R(Q)}(x;\lam,
\pi \pm \phi) \; ,
\ee
which reflect directly into the MDFs,
\br
\lb{e22}
Q(q;\lam,\phi) &=& R(q;\lam^{-1},\phi) = R(q;\lam,\pi \pm \phi) \; , \nn \\
R(p;\lam,\phi) &=& Q(p;\lam^{-1},\phi) = Q(p;\lam,\pi \pm \phi) \; .
\er
Consequently, the calculation of $Q(q;\lam,\phi)$ is sufficient for 
determining the function $R(p;\lam,\phi)$, and vice-versa.

Now we analyze the structure of the kernel (\ref{e6}), which can be factorized 
as
\be
\lb{e23}
K(\xi,\eta;\lam,\phi) = k_{Q}(\xi;\lam,\phi) k_{R}(\eta;\lam,\phi) 
k_{C}(\xi,\eta;\lam,\phi) \; ,
\ee
where the first two factors on the RHS (reduced kernels) `propagate' the 
initial HF in an independent way, {\em i.e.}, if in the `initial' $(\lam = 1)$ 
HF the phase-space variables are not correlated, they will remain as such for 
any other value of $\lam$. The factor
\be
\lb{e24}
k_{C}(\xi,\eta;\lam,\phi) = \exp \lbk - \lpar \frac{\lam^{2} - 1}{4 \lam} 
\sin \phi \rpar \xi \eta \rbk 
\ee
introduces additional (or new) correlations into an `initial' HF when $\phi 
\neq n \pi$, with $n \in {\rm I\!N}$. Otherwise, we obtain $k_{C}=1$ and 
$K(\xi,\eta;\lam,n \pi) = k_{Q}(\xi;\lam,n \pi) k_{R}(\eta;\lam,n \pi)$,
one reduced kernel for each variable.

As a consequence of the factorization (\ref{e23}) it is interesting to rewrite 
the CDF $C(p,q;\lam,\phi)$ as a sum of two terms,
\be
\lb{e25}
C(p,q;\lam,\phi) = C^{(1)}(p,q;\lam,\phi) + C^{(2)}(p,q;\lam,\phi) \; ,
\ee
defined as
\be
\lb{e26}
C^{(1)}(p,q;\lam,\phi) = \int_{- \infty}^{\infty} \frac{d\xi d\eta}{2 \pi} \; 
e^{\im (\eta p - \xi q)} K(\xi,\eta;\lam,\phi) \widetilde{C}(\xi,\eta) 
\ee
and
\be
\lb{e27}
C^{(2)}(p,q;\lam,\phi) = \int_{- \infty}^{\infty} \frac{d\xi d\eta}{2 \pi} \; 
e^{\im (\eta p - \xi q)} k_{Q}(\xi;\lam,\phi) k_{R}(\eta;\lam,\phi) \lbk k_{C}
(\xi,\eta;\lam,\phi) - 1 \rbk \widetilde{R}(\eta) \widetilde{Q}(\xi) \; . 
\ee
Here we assume that the HF contains `initial' correlations [see equation 
(\ref{e12})] with its FT being $\widetilde{P}(\xi,\eta) = \widetilde{R}(\eta) 
\widetilde{Q}(\xi) + \widetilde{C}(\xi,\eta)$, and that `propagation' of 
correlations originates from two sources. The first, the RHS of equation 
(\ref{e26}), is responsible for the `propagation' of squeezing into the 
`initial' correlations $\widetilde{C}(\xi,\eta)$. In the second, equation 
(\ref{e27}), `propagation' occurs only for $\phi \neq n \pi$ when additional 
correlations are created into the `initial' FT of the uncorrelated part of the 
HF, $\widetilde{R}(\eta) \widetilde{Q}(\xi)$. In appendix B, the correlations 
$C^{(1)}$ and $C^{(2)}$ are obtained for the Fock states in the SSR.

\section{Fock states in the squeezed states representation}

The density operator $\ro_{n} = |n \rg \lg n|$ mapped in the coherent states 
representation yields a Poisson distribution \cite{s11}
\be
\lb{e28}
P_{n}(p,q) = | \lg pq | n \rg |^{2} = \frac{1}{n!} \lpar \frac{p^{2} + q^{2}}
{2} \rpar^{n} \exp \lpar - \frac{p^{2} + q^{2}}{2} \rpar \; ,
\ee
with $n = 0,1,2,\ldots$. So, the respective Fourier transform
\be
\lb{e29}
\widetilde{P}_{n}(\xi,\eta) = \frac{(-1)^{n}}{2^{2n} n!} \sum_{k=0}^{n} 
\frac{n!}{k! (n-k)!} \; {\cal H}_{2k} \lpar \frac{\xi}{\sqrt{2}} \rpar 
{\cal H}_{2(n-k)} \lpar \frac{\eta}{\sqrt{2}} \rpar \exp \lpar -
\frac{\xi^{2} + \eta^{2}}{2} \rpar \; ,
\ee
where ${\cal H}_{m}(x)$ is the Hermite polynomial, represents an initial step
for calculating the Husimi function $P_{n}(p,q;\lam,\phi)$ in the squeezed 
states representation. In fact, substituting equation (\ref{e29}) into
(\ref{e5}) and evaluating the integrations over $\xi$ and $\eta$
\cite[\S 7.374-8]{s12}, we get
\be
\lb{e30}
P_{n}(p,q;\lam,\phi) = \frac{2 \sqrt{\lam}}{\lam + 1} \lpar \frac{\lam - 1}
{\lam + 1} \rpar^{n} \frac{1}{2^{n} n!} \left| {\cal H}_{n} \lpar 
\frac{\lam q_{r} + \im p_{r}}{\sqrt{\lam^{2} - 1}} \rpar \right|^{2} \exp 
\lpar - \frac{\lam q_{r}^{2} + p_{r}^{2}}{\lam + 1} \rpar \; ,
\ee
where $q_{r}$ and $p_{r}$ are the rotated variables defined in section II. 
This expression was inittialy obtained in \cite{s13}, and later used by
Schleich {\em et al}. \cite{s1} in the oscillatory behavior study of the
distribution $P_{n}(p,q;\lam,0)$. Now, using the mathematical relation
\cite{s14}
\bd
\left| {\cal H}_{n}(z) \right|^{2} = 2^{n} n! \sum_{k=0}^{n} \; (-1)^{k} 
{\cal L}_{k}^{(-1/2)}(2 x^{2}) \; {\cal L}_{n-k}^{(-1/2)}(-2 y^{2}) \qquad 
(z = x + \im y) \; ,
\ed
in which ${\cal L}_{m}^{(\alf)}(x)$ is the associated Laguerre polynomial,
equation (\ref{e30}) can be written in an equivalent form,
\be
\lb{e31}
P_{n}(p,q;\lam,\phi) = \frac{2 \sqrt{\lam}}{\lam + 1} \lpar \frac{\lam - 1}
{\lam + 1} \rpar^{n} \sum_{k=0}^{n} \; (-1)^{k} {\cal L}_{k}^{(-1/2)} \lpar
\frac{2 \lam^{2} q_{r}^{2}}{\lam^{2} - 1} \rpar {\cal L}_{n-k}^{(-1/2)}
\lpar - \frac{2 p_{r}^{2}}{\lam^{2} - 1} \rpar \exp \lpar - \frac{\lam 
q_{r}^{2} + p_{r}^{2}}{\lam + 1} \rpar \; . 
\ee

The Husimi function $Q_{n}(q)$ is obtained with the help of equation
(\ref{e28}), {\em i.e.},
\be
\lb{e32}
Q_{n}(q) = \int_{- \infty}^{\infty} \frac{dp}{\sqrt{2 \pi}} \; P_{n}(p,q) =
\exp \lpar - \frac{q^{2}}{2} \rpar \sum_{k=0}^{n} {\cal L}_{n-k}^{(-1/2)}(0) \; 
\frac{q^{2k}}{2^{k} k!} \; , 
\ee
whose Fourier transform is given by
\be
\lb{e33}
\widetilde{Q}_{n}(\xi) = \sum_{k=0}^{n} {\cal L}_{n-k}^{(-1/2)}(0) \; 
{\cal L}_{k}^{(-1/2)} \lpar \frac{\xi^{2}}{2} \rpar \exp \lpar - \frac{\xi^{2}}
{2} \rpar \; . 
\ee
Substituting this result into equation (\ref{e15}) and doing the integration
with respect to $\xi$, we get
\br
\lb{e34}
Q_{n}(q;\lam,\phi) &=& \sqrt{\frac{2 \lam}{(\lam + 1) [ \cos^{2} (\phi/2) + 
\lam \sin^{2} (\phi/2)]}} \; \sum_{k=0}^{n} \; (-1)^{k} {\cal L}_{n-k}^{(-1/2)}
(0) \lbk \frac{\lam - 1}{\lam + 1} \frac{\cos^{2} (\phi/2) - \lam \sin^{2} 
(\phi/2)}{\cos^{2} (\phi/2) + \lam \sin^{2} (\phi/2)} \rbk^{k} \nn \\ 
& & \times {\cal L}_{k}^{(-1/2)} \lbk \frac{2 \lam^{2} q^{2}}{(\lam^{2}-1)[ 
\cos^{4} (\phi/2) - \lam^{2} \sin^{4} (\phi/2)]} \rbk \exp \lbk - \frac{\lam 
q^{2}}{(\lam + 1) [ \cos^{2} (\phi/2) + \lam \sin^{2} (\phi/2)]} \rbk \; .
\er
In order to obtain the MDF $R_{n}(p;\lam,\phi)$, we only need the symmetry
properties (\ref{e22}),
\br
\lb{e35}
R_{n}(p;\lam,\phi) &=& \sqrt{\frac{2 \lam}{(\lam + 1) [ \lam \cos^{2} (\phi/2) 
+ \sin^{2} (\phi/2)]}} \; \sum_{k=0}^{n} \; {\cal L}_{k}^{(-1/2)}(0) \lbk 
\frac{\lam - 1}{\lam + 1} \frac{\lam \cos^{2} (\phi/2) - \sin^{2} (\phi/2)}
{\lam \cos^{2} (\phi/2) + \sin^{2} (\phi/2)} \rbk^{n-k} \nn \\
& & \times {\cal L}_{n-k}^{(-1/2)} \lbk - \frac{2 \lam^{2} p^{2}}{(\lam^{2}-1)
[ \lam^{2} \cos^{4} (\phi/2) - \sin^{4} (\phi/2)]} \rbk \exp \lbk - \frac{\lam 
p^{2}}{(\lam + 1) [ \lam \cos^{2} (\phi/2) + \sin^{2} (\phi/2)]} \rbk \; .
\er

Figures 1(a)-(d) show the three-dimensional plots of $P_{n}(p,q;\lam,\phi)$ versus $n$ 
and $\phi$, for $\lam = 21,201,1/21,1/201$, respectively. The plane $\phi =0$ in figure 
1(a) corresponds to the oscillations pointed out in \cite{s1}, which depend strongly on 
$\phi$, showing a periodicity of $\pi$. Now, for $\lam=201$ [figure 1(b)] we observe the 
occurence of rich structures, although the beats pointed out in \cite{s4} can not be 
perceived. In fact, they are revealed in figures 2 and 3. Figures 2(a)-(f) show the plots 
of $P_{n}$ versus $n$ for $\phi =85^{o},...,90^{o}$ and $\lam = 201$, where the beat 
structure becomes evident; however, it disappears at angles close to $90^{o}$. Following 
the arguments presented in \cite{s4} and corroborated by Mandal \cite{s7}, this beat 
structure is a consequence of the quantum interference in phase space. Figures 3(a)-(f) show 
the plots of $C_{n}$ versus $n$ for the same parameters used in figures 2(a)-(f), where the 
beat structure is present again. This fact connects the {\em correlations} and 
{\em interference} effects in phase space, and provides further insights to the phenomenon. 
Moreover, we observe a similar kind of plots for $\lam =1/21$ and $\lam = 1/201$, figures 
1(c) and 1(d), respectively, where now they are shifted by $\pi/2$.

\section{Summary and conclusions}

We have considered the Husimi function $P(p,q;\lam,\phi)$ with emphasis on the
marginal and correlation distribution functions, showing that all three
satisfy the pseudo-diffusion equations if one considers that the squeezing
parameter $\lam$ plays the role of a time. The solution, obtained from the
Fourier transform method, permits calculating $P(p,q;\lam,\phi)$, given an
`initial' Husimi function $P(p,q)$ with a kernel $K(\xi,\eta;\lam,\phi)$
responsible by the propagation of squeezing. The decomposition of the kernel
in three factors, equation (\ref{e23}), permits writing the CDF as a sum of
two terms, having different interpretations: the first term, equation \rf{e26},
is the propagation of `initial' correlations contained in $P(p,q)$; whereas
the second term, equation \rf{e27}, is responsible for introducing additional 
correlations into the uncorrelated `initial' product of the MDFs $Q(q) R(p)$.

Finally we remind that the formal procedure employed throughout this paper is
advantageous if compared with the direct and lengthy calculation exposed in
appendix A. In the specific case of the number state, the decomposition of the
CDF into two terms should permit to investigate more thouroughly the origin of
beats. Although having attained a formal expression for both (see appendix B), 
the numerical calculation presents difficulties due to its complexity.

Multimode-squeezed-states representation can also be considered within the
present formalism. In particular, M. Selvadoray {\em et al}. \cite{s15}
studied the two-mode-squeezed-state photon distribution. Again they verified
the presence of beats. In this case, the correlation distribution function
plays a crucial role in the understanding of this effect.

\section*{Acknowledgments}

MAM acknowledges financial support from FAPESP, S\~ao Paulo, project number 97/14551-4. 
SSM acknowledges financial support from CNPq, Brasil. This work has also been partially 
supported by Conv\^enio FINEP/PRONEX Grant number 41/96/0935/00.

\appendix
\section {MDFs for the number states by direct integration}

The usual procedure to calculate the MDFs for the number states consists in 
the integration of equation (\ref{e30}) with respect the variables $p$ or $q$, 
respectively,
\br
\lb{A1}
Q_{n}(q;\lam,\phi) &=& \int_{- \infty}^{\infty} \frac{dp}{\sqrt{2 \pi}} \; 
P_{n}(p,q;\lam,\phi) \; , \\
\lb{A2}
R_{n}(p;\lam,\phi) &=& \int_{- \infty}^{\infty} \frac{dq}{\sqrt{2 \pi}} \; 
P_{n}(p,q;\lam,\phi) \; .
\er
We calculate (\ref{A1}) by direct integration and present some properties 
inherent to the marginal distributions.

Consider initially the integral representation of the Hermite polynomial 
\cite{s16}
\bd
{\cal H}_{n}(z) = (- 2 \im)^{n} e^{z^{2}} \int_{- \infty}^{\infty} \frac{du}
{\sqrt{\pi}} \; u^{n} \; e^{- u^{2} + 2 \im u z} \; , 
\ed
which permits to write (\ref{e30}) in a more convenient form:
\br
P_{n}(p,q;\lam,\phi) &=& \frac{2 \sqrt{\lam}}{\lam + 1} \frac{(-2)^{n}}{n!} 
\lpar \frac{\lam - 1}{\lam + 1} \rpar^{n} \exp \lbr - \lbk \frac{\sin^{2} 
(\phi/2) - \lam \cos^{2} (\phi/2)}{\lam - 1} \rbk q^{2} \rbr \nn \\
& & \times \exp \lbr - \lbk \frac{\cos^{2} (\phi/2) - \lam \sin^{2} (\phi/2)}
{\lam - 1} \rbk p^{2} + \lpar \frac{\lam + 1}{\lam -1} \; \sin \phi \rpar pq 
\rbr \nn \\
& & \times \int_{- \infty}^{\infty} \frac{du dv}{\pi} \; (uv)^{n} \exp \lbr 
- (u^{2} + v^{2}) + \frac{2}{\sqrt{\lam^{2} - 1}} \lbk (u-v) \sin (\phi/2) + 
\im \lam (u+v) \cos (\phi/2) \rbk q \rbr \nn \\
& & \times \exp \lbr - \frac{2}{\sqrt{\lam^{2} - 1}} \lbk (u-v) \cos (\phi/2) 
- \im \lam (u+v) \sin (\phi/2) \rbk p \rbr \; . \lb{A3}
\er
Substituting (\ref{A3}) into (\ref{A1}) and integrating with respect to $p$, 
we obtain
\br
Q_{n}(q;\lam,\phi) &=& \sqrt{2 \alf^{2}} \; \frac{(-2)^{n}}{n!} \lpar 
\frac{\alf}{\beta} \rpar^{n} e^{- (\alf q)^{2}} \int_{- \infty}^{\infty} 
\frac{dy}{\sqrt{\pi}} \; y^{n} e^{- (y - \im \alf q)^{2}} \nn \\
& & \times \int_{- \infty}^{\infty} \frac{dx}{\sqrt{\pi}} \; x^{n} \exp \lbr  
- \lbk x - \im \lpar \beta q + \im \; \frac{\beta}{2 \alf} \lpar 1 - 
\frac{\alf^{2}}{\beta^{2}} \rpar y \rpar \rbk^{2} \rbr \lb{A4}
\er
for values of the squeeze parameter in the intervals $1 < \lam < \cot^{2}
(\phi/2)$ or $\cot^{2} (\phi/2) < \lam < 1$, where
\bd
\alf = \sqrt{\frac{\lam}{(\lam + 1) \lbk \cos^{2} (\phi/2) + \lam \sin^{2} 
(\phi/2) \rbk}} \qquad \mbox{and} \qquad \beta = \sqrt{\frac{\lam}{(\lam - 1) 
\lbk \cos^{2} (\phi/2) - \lam \sin^{2} (\phi/2) \rbk}} \; .
\ed
Integration over the variable $x$ leads to \cite[\S 3.462-4]{s12}
\be
Q_{n}(q;\lam,\phi) = \sqrt{2 \alf^{2}} \; \frac{(- \im)^{n}}{n!} \lpar 
\frac{\alf}{\beta} \rpar^{n} e^{- (\alf q)^{2}} \int_{- \infty}^{\infty} 
\frac{dy}{\sqrt{\pi}} \; y^{n} \; {\cal H}_{n} \lbk \beta q + \im \; 
\frac{\beta}{2 \alf} \lpar 1 - \frac{\alf^{2}}{\beta^{2}} \rpar y \rbk 
e^{- (y - \im \alf q)^{2}} \; . \lb{A5}
\ee
Now, using relation \cite{s14}
\bd
{\cal H}_{n}(z+w) = \sum_{k=0}^{n} {\cal L}_{n-k}^{(k)}(0) \; (2w)^{n-k} 
{\cal H}_{k}(z) 
\ed
for the Hermite polynomial present in (\ref{A5}),
\bd
{\cal H}_{n} \lbk \beta q + \im \; \frac{\beta}{2 \alf} \lpar 1 - 
\frac{\alf^{2}}{\beta^{2}} \rpar y \rbk = \sum_{k=0}^{n} {\cal L}_{n-k}^{(k)}(0) 
\lbk \im \; \frac{\beta}{\alf} \lpar 1 - \frac{\alf^{2}}{\beta^{2}} \rpar y 
\rbk^{n-k} {\cal H}_{k} (\beta q) \; ,
\ed
we get
\be
Q_{n}(q;\lam,\phi) = \frac{\sqrt{2 \alf^{2}}}{2^{n} n!} \lpar \frac{\alf}
{\beta} \rpar^{n} e^{- (\alf q)^{2}} \sum_{k=0}^{n} {\cal L}_{n-k}^{(k)}(0) 
\lbk \frac{\beta}{2 \alf} \lpar \frac{\alf^{2}}{\beta^{2}} - 1 \rpar \rbk^{n-k} 
{\cal H}_{k} (\beta q) {\cal H}_{2n-k}(\alf q) \; . \lb{A6}
\ee
Note that (\ref{A6}) can be written in a compact form and equivalent to 
equation (\ref{e34}), {\em i.e.}, in terms of the associated Laguerre 
polynomial. For this purpose it is necessary to verify the equality
\be
\frac{1}{2^{2n} n!} \sum_{s=0}^{n} {\cal L}_{n-s}^{(s)}(0) \lpar \frac{2 \alf}
{\beta} \rpar^{s} \lpar \frac{\alf^{2}}{\beta^{2}} - 1 \rpar^{n-s} 
{\cal H}_{s}(\beta q) {\cal H}_{2n-s}(\alf q) = \sum_{k=0}^{n} c_{nk}(\alf,
\beta) \; {\cal L}_{k}^{(-1/2)} \lbk 2 (\alf \beta q)^{2} \rbk \lb{A7}
\ee
and to determine the coefficients $c_{nk}(\alf,\beta)$.

The relation established by Bailey \cite{s17} for the product of Hermite polynomials,
\br
{\cal H}_{m}(ax) {\cal H}_{l}(bx) &=& \sum_{j=0}^{\lbk \frac{m+l}{2} \rbk} 
(-1)^{j} \frac{m!}{j! (m-2j)!} \frac{a^{m-2j} b^{2j-l}}{\lpar \sqrt{a^{2} + 
b^{2}} \; \rpar^{m-l}} \; _{2}{\cal F}_{1} \lpar m+1, -l; m-2j+1; \frac{a^{2}}
{a^{2} + b^{2}} \rpar \nn \\
& & \times \; {\cal H}_{m+l-2j} \lpar \sqrt{a^{2} + b^{2}} \; x \rpar \nn
\er
where $_{2}{\cal F}_{1}(a_{1},a_{2};a_{3};z)$ is the hypergeometric function, 
permits to verify the equality (\ref{A7}) through
\br
{\cal H}_{s}(\beta q) {\cal H}_{2n-s}(\alf q) &=& \frac{(-1)^{n} (2n-s)!}
{2^{n-s} (\alf \beta)^{2n-s}} \sum_{k=0}^{n} \frac{k!}{(2k-s)! (n-k)!} \; 
(2 \alf)^{2k} \beta^{2(n-k)} \nn \\
& & \times \; _{2}{\cal F}_{1} \lpar 2n-s+1, -s; 2k-s+1; \frac{1}{2 \beta^{2}} 
\rpar {\cal L}_{k}^{(-1/2)} \lbk 2 (\alf \beta q)^{2} \rbk \; , \lb{A8}
\er
and to determine the coefficients $c_{nk}(\alf,\beta)$,
\br
c_{nk}(\alf,\beta) &=& \frac{(-1)^{n}}{2^{2n}} \frac{k!}{(n-k)!} 
\frac{(2 \alf)^{2k} \beta^{2(n-k)}}{(2 \alf^{2} \beta^{2})^{n}} \sum_{s=0}^{n} 
\frac{(2n-s)!}{s! (n-s)! (2k-s)!} \; (2 \alf)^{2s} \lpar \frac{\alf^{2}}
{\beta^{2}} - 1 \rpar^{n-s} \nn \\
& & \times \; _{2}{\cal F}_{1} \lpar 2n-s+1, -s; 2k-s+1; \frac{1}{2 \beta^{2}} 
\rpar \; . \lb{A9}
\er
In fact, the sum present in (\ref{A9}) can be performed since we used the 
following relation for the hypergeometric function \cite{s14}:
\be
_{2}{\cal F}_{1} \lpar 2n-s+1, -s; 2k-s+1; \frac{1}{2 \beta^{2}} \rpar = 
\frac{s! (2k-s)!}{(2n-s)!} \sum_{l=0}^{s} (-1)^{l} \frac{(2n-s+l)!}{(s-l)! 
(2k-s+l)!} \frac{(2 \beta^{2})^{-l}}{l!} \; . \lb{A10}
\ee
Then, substituting (\ref{A10}) into (\ref{A9}) and calculating the sums, we
obtain a simple expression for the coefficients
\be
c_{nk}(\alf,\beta) = (-1)^{k} \; \frac{[2(n-k)-1]!!}{[2(n-k)]!!} \lpar 
\frac{\alf}{\beta} \rpar^{2k} = (-1)^{k} {\cal L}_{n-k}^{(-1/2)}(0) \lpar 
\frac{\alf}{\beta} \rpar^{2k} \; . \lb{A11}
\ee
So, the marginal distribution (\ref{A6}) can be expressed such as equation
(\ref{e34}). Adopting analogous procedure for equation (\ref{A2}) we obtain
(\ref{e35}), however the values of squeeze parameter are restricted in the
intervals $\tan^{2} (\phi/2) > \lam > 1$ or $\tan^{2} (\phi/2) < \lam < 1$.

In addition to the symmetry relations established in section III [see equation
(\ref{e22})], the MDFs exhibit the following properties:
\br
(i) \; \lim_{\lam \rightarrow \infty} Q_{n}(q;\lam,0) &=& | \Psi_{n}(q) |^{2} 
\qquad \mbox{and} \qquad \lim_{\lam \rightarrow 0} R_{n}(p;\lam,0) = | 
\Phi_{n}(p) |^{2} \; , \lb{A12} \\
(ii) \; \sum_{n=0}^{\infty} Q_{n}(q;\lam,\phi) &=& \sum_{n=0}^{\infty} R_{n}
(p;\lam,\phi) = \sum_{n=0}^{\infty} {\cal L}_{n}^{(-1/2)}(0) \rightarrow \infty 
\; . \lb{A13}
\er
The first property does not characterize the Husimi function as a probability 
distribution but only emphasizes the character of quasiprobability
distribution: limit values of $\lam$ recover the squared moduli of
wavefunctions \cite{s10,s18}. With respect to the second property, it is a
direct consequence of the scalar product for the squeezed states, {\em i.e.}, 
$\sum_{n=0}^{\infty} P_{n}(p,q;\lam,\phi) = 1$. In fact, the divergence is a 
consequence of the integration step of this relation over the variables $p$ or 
$q$. Now, considering the normalization of equation (\ref{e30}),
$\int_{- \infty}^{\infty} \frac{dp dq}{2 \pi} \; P_{n}(p,q;\lam,\phi) = 1$,
we obtain the third property for MPDFs:
\be
(iii) \; \int_{- \infty}^{\infty} \frac{dq}{\sqrt{2 \pi}} \; Q_{n}(q;\lam,
\phi) = 1 \qquad \mbox{and} \qquad \int_{- \infty}^{\infty} \frac{dp}{\sqrt{2 
\pi}} \; R_{n}(p;\lam,\phi) = 1 \; . \lb{A14}
\ee

\section{The components of the correlation distribution function}

The components of the correlation distribution function $C(p,q;\lam,\phi)$ 
can also be written as
\br
C^{(1)}(p,q;\lam,\phi) &=& P(p,q;\lam,\phi) - C^{(3)}(p,q;\lam,\phi) \; ,
\lb{B1} 
\\
C^{(2)}(p,q;\lam,\phi) &=& C^{(3)}(p,q;\lam,\phi) - Q(q;\lam,\phi) R(p;\lam,
\phi) \; , \lb{B2}
\er
with
\be
C^{(3)}(p,q;\lam,\phi) = \int_{- \infty}^{\infty} \frac{d\xi d\eta}{2 \pi} 
\; e^{\im (\eta p - \xi q)} K(\xi,\eta;\lam,\phi) \; \widetilde{Q}(\xi;1) 
\widetilde{R}(\eta;1) \; . \lb{B3}
\ee
Present in both equations (\ref{B1}) and (\ref{B2}), the term $C^{(3)}$ is 
responsible for introducing correlations into the marginal distributions 
product through the relation
\be
C^{(3)}(p,q;\lam,\phi) = \exp \lbk \lpar \frac{\lam^{2} - 1}{4 \lam} \sin 
\phi \rpar \frac{\partial^{2}}{\partial p \partial q} \rbk Q(q;\lam,\phi) 
R(p;\lam,\phi) \; , \lb{B4}
\ee
where we used the result:
\bd
e^{\im (\eta p - \xi q)} K(\xi,\eta;\lam,\phi) = \exp \lbk \lpar 
\frac{\lam^{2} - 1}{4 \lam} \sin \phi \rpar \frac{\partial^{2}}{\partial p 
\partial q} \rbk e^{\im (\eta p - \xi q)} k_{Q}(\xi;\lam,\phi) k_{R}(\eta;
\lam,\phi) \; .
\ed
For $\phi=0$, equations (\ref{B1})-(\ref{B3}) simplify to
\br
C^{(1)}(p,q;\lam,0) &=& C(p,q;\lam,0) \; , \nn \\
C^{(2)}(p,q;\lam,0) &=& 0 \; , \nn \\
C^{(3)}(p,q;\lam,0) &=& Q(q;\lam,0) R(p;\lam,0) \; . \nn
\er
In this appendix we calculate the component $C^{(3)}$ for the number states
and, consequently, the components $C^{(1)}$ and $C^{(2)}$ can be totally
determined.

In order to simplify the calculations, let us initially consider the marginal 
and correlation distribution functions expressed in terms of the parameters 
$\alf_{1}$, $\alf_{2}$ and $\alf_{3}$:
\br
Q_{n}(q;\lam,\phi) &=& \sqrt{2 \alf_{1}^{2}} \; e^{- (\alf_{1} q)^{2}} 
\sum_{k=0}^{n} {\cal L}_{n-k}^{(k)}(0) \; (-2 \alf_{1}^{2})^{k} \; 
{\cal L}_{k}^{(-1/2)} \lbk (\alf_{1} q)^{2} \rbk \; , \lb{B5} \\
R_{n}(p;\lam,\phi) &=& \sqrt{2 \alf_{2}^{2}} \; e^{- (\alf_{2} p)^{2}} 
\sum_{k=0}^{n} {\cal L}_{n-k}^{(k)}(0) \; (-2 \alf_{2}^{2})^{k} \; 
{\cal L}_{k}^{(-1/2)} \lbk (\alf_{2} p)^{2} \rbk \; , \lb{B6} \\
C_{n}^{(3)}(p,q;\lam,\phi) &=& \sum_{l=0}^{\infty} \frac{\alf_{3}^{l}}{l!} 
\frac{\partial^{2l}}{\partial p^{l} \partial q^{l}} \; Q_{n}(q;\lam,\phi) 
R_{n}(p;\lam,\phi) \; , \lb{B7}
\er
with
\br
\alf_{1} &=& \sqrt{\frac{\lam}{(\lam + 1) \lbk \cos^{2} (\phi/2) + \lam 
\sin^{2} (\phi/2) \rbk}} \; , \nn \\
\alf_{2} &=& \sqrt{\frac{\lam}{(\lam + 1) \lbk \lam \cos^{2} (\phi/2) + 
\sin^{2} (\phi/2) \rbk}} \; , \nn \\
\alf_{3} &=& \frac{\lam^{2} - 1}{4 \lam} \; \sin \phi \; , \nn
\er
and $\alf_{1}^{2} + \alf_{2}^{2} = 2 \alf_{1}^{2} \alf_{2}^{2}$. Expressions 
(\ref{B5}) and (\ref{B6}) are the alternative way of writing the marginal 
distributions (\ref{e34}) and (\ref{e35}), respectively, since we used the 
properties of the associated Laguerre polynomials. Moreover, using the 
relation
\bd
\frac{\partial^{l}}{\partial x^{l}} \lbr e^{-(ax)^{2}} {\cal L}_{k}^{(-1/2)}
[(ax)^{2}] \rbr = \frac{(-1)^{k}}{2^{2k} k!} \; (-a)^{l} e^{-(ax)^{2}} 
{\cal H}_{l+2k}(ax) \; ,
\ed
those expressions permit to calculate the component $C_{n}^{(3)}$, given the 
result:
\br
C_{n}^{(3)}(p,q;\lam,\phi) &=& 2 \alf_{1} \alf_{2} \; e^{- [ (\alf_{1} q)^{2} 
+ (\alf_{2} p)^{2} ]} \sum_{k=0}^{n} {\cal L}_{n-k}^{(k)}(0) \; \frac{\alf_{1}^{2k}}
{2^{k} k!} \sum_{m=0}^{n} {\cal L}_{n-m}^{(m)}(0) \; \frac{\alf_{2}^{2m}}{2^{m} m!} 
\nn \\
& & \times \sum_{l=0}^{\infty} \frac{( \alf_{1} \alf_{2} \alf_{3} )^{l}}{l!} 
\; {\cal H}_{l+2k}(\alf_{1} q) {\cal H}_{l+2m}(\alf_{2} p) \; . \lb{B8} 
\er
The infinity sum present in (\ref{B8}) can be done by using the formula 
\cite[\S 5.12.2.1]{s19}
\br
\sum_{k=0}^{\infty} \frac{t^{k}}{k!} \; {\cal H}_{k+m}(x) {\cal H}_{k+n}(y) 
&=& \frac{1}{\sqrt{(1-4t^{2})^{m+n+1}}} \; \exp \lbk \frac{4txy - 4t^{2} 
(x^{2} + y^{2})}{1 - 4t^{2}} \rbk \sum_{r=0}^{ \{ m,n \} } r! \; {\cal L}_{m-r}^{(r)}(0) 
\; {\cal L}_{n-r}^{(r)}(0) \; (4t)^{r} \nn \\
& & \times \; {\cal H}_{m-r} \lpar \frac{x-2ty}{\sqrt{1-4t^{2}}} \rpar 
{\cal H}_{n-r} \lpar \frac{y-2tx}{\sqrt{1-4t^{2}}} \rpar \qquad |t| < 1/2 \; ,
\nn
\er
where $\{ m,n \}$ stands for the minor of $m$ and $n$, which leads to 
\br
C_{n}^{(3)}(p,q;\lam,\phi) &=& \frac{2 \alf_{1} \alf_{2}}{\sqrt{1 - ( 2 
\alf_{1} \alf_{2} \alf_{3})^{2}}} \; \exp \lbk - \frac{\alf_{1}^{2} q^{2} - 4 
\alf_{1}^{2} \alf_{2}^{2} \alf_{3} pq + \alf_{2}^{2} p^{2}}{1 - (2 \alf_{1} 
\alf_{2} \alf_{3})^{2}} \rbk \sum_{k=0}^{n} \frac{{\cal L}_{n-k}^{(k)}(0)}
{2^{k} k!} \lbk \frac{\alf_{1}^{2}}{1 - (2 \alf_{1} \alf_{2} \alf_{3})^{2}} 
\rbk^{k} \nn \\
& & \times \sum_{m=0}^{n} \frac{{\cal L}_{n-m}^{(m)}(0)}{2^{m} m!} \lbk 
\frac{\alf_{2}^{2}}{1 - (2 \alf_{1} \alf_{2} \alf_{3})^{2}} \rbk^{m} \; 
\sum_{r=0}^{ \{ 2k,2m \} } r! \; {\cal L}_{2k-r}^{(r)}(0) \; {\cal L}_{2m-r}^{(r)}(0)
\; ( 4 \alf_{1} \alf_{2} \alf_{3})^{r} \nn \\
& & \times \; {\cal H}_{2k-r} \lpar \frac{\alf_{1} q - 2 \alf_{1} \alf_{2}^{2} 
\alf_{3} p}{\sqrt{1 - (2 \alf_{1} \alf_{2} \alf_{3})^{2}}} \rpar 
{\cal H}_{2m-r} \lpar \frac{\alf_{2} p - 2 \alf_{1}^{2} \alf_{2} \alf_{3} q}
{\sqrt{1 - (2 \alf_{1} \alf_{2} \alf_{3})^{2}}} \rpar \; . \lb{B9}
\er



\begin{figure}
\caption{Plots of $P_{n}(p,q;\lam,\phi)$ vs. $n$ and $\phi$ with
$p^{2} + q^{2} = 98$. Figures 1(a)-(d) correspond to $\lam = 21,201,1/21$
and $1/201$, respectively.}
\end{figure}
\begin{figure}
\caption{Plots of $P_{n}(p,q;\lam,\phi)$ vs. $n$ with 
$\phi = 85^{o}, \ldots, 90^{o}$ and $\lam = 201$. The phase-space variables
are transformed into $q = 7 \sqrt{2} \cos \th $ and $p = 7 \sqrt{2} \sin \th$,
which leads to $q_{r} = 7 \sqrt{2} \cos (\th - \phi/2)$ and
$p_{r} = 7 \sqrt{2} \sin (\th - \phi/2)$. For mapping the parameters used by
Dutta {\em et al.}, we fixed $\th = 3 \phi /2$.}
\end{figure}
\begin{figure}
\caption{Plots of $C_{n}(p,q;\lam,\phi)$ vs. $n$ for the same parameters
set used in figure 2, where we see the presence of beats again. This
fact corroborates the phase-space interference concept since
{\em correlations} and {\em interference} effects are closely connected.}
\end{figure}
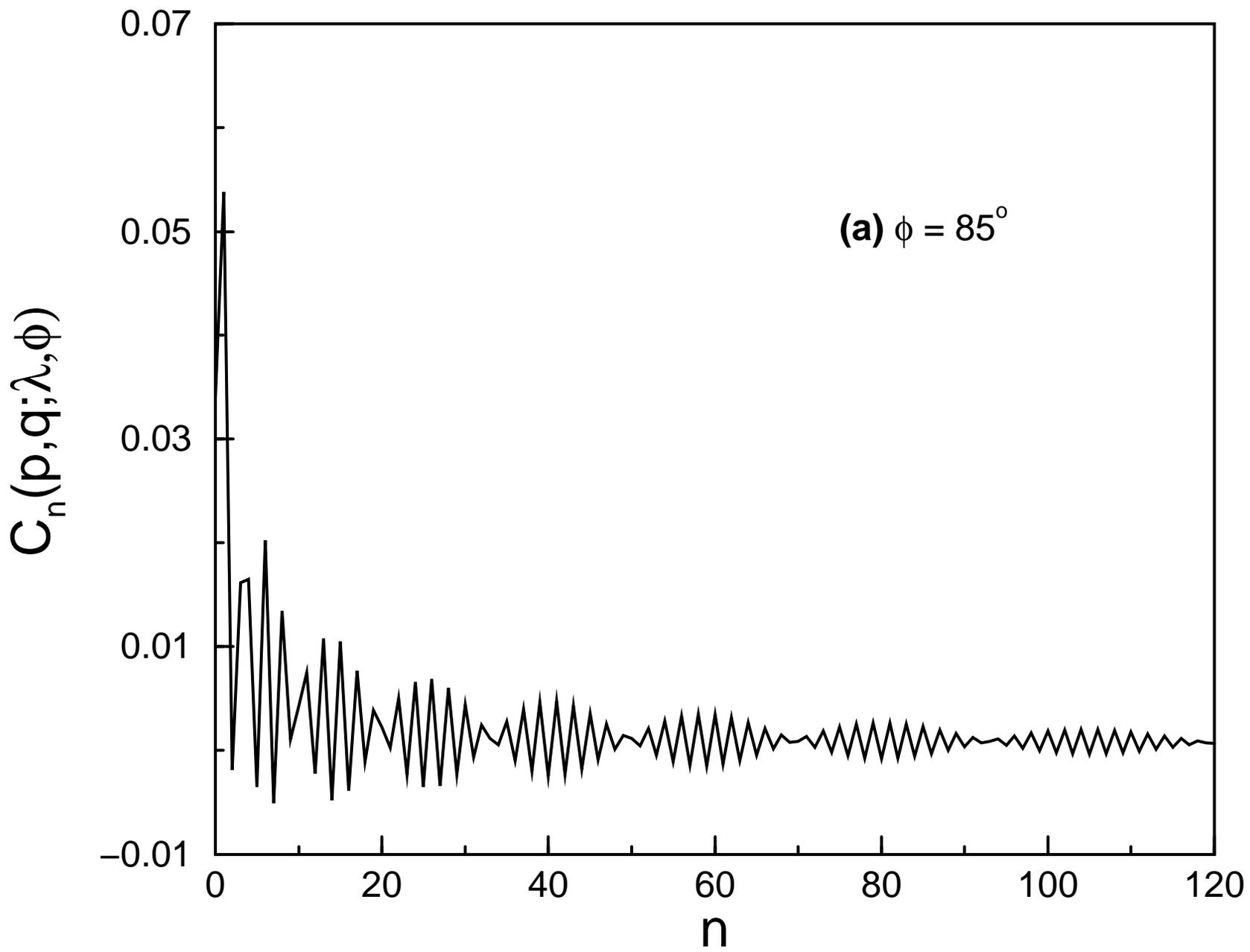
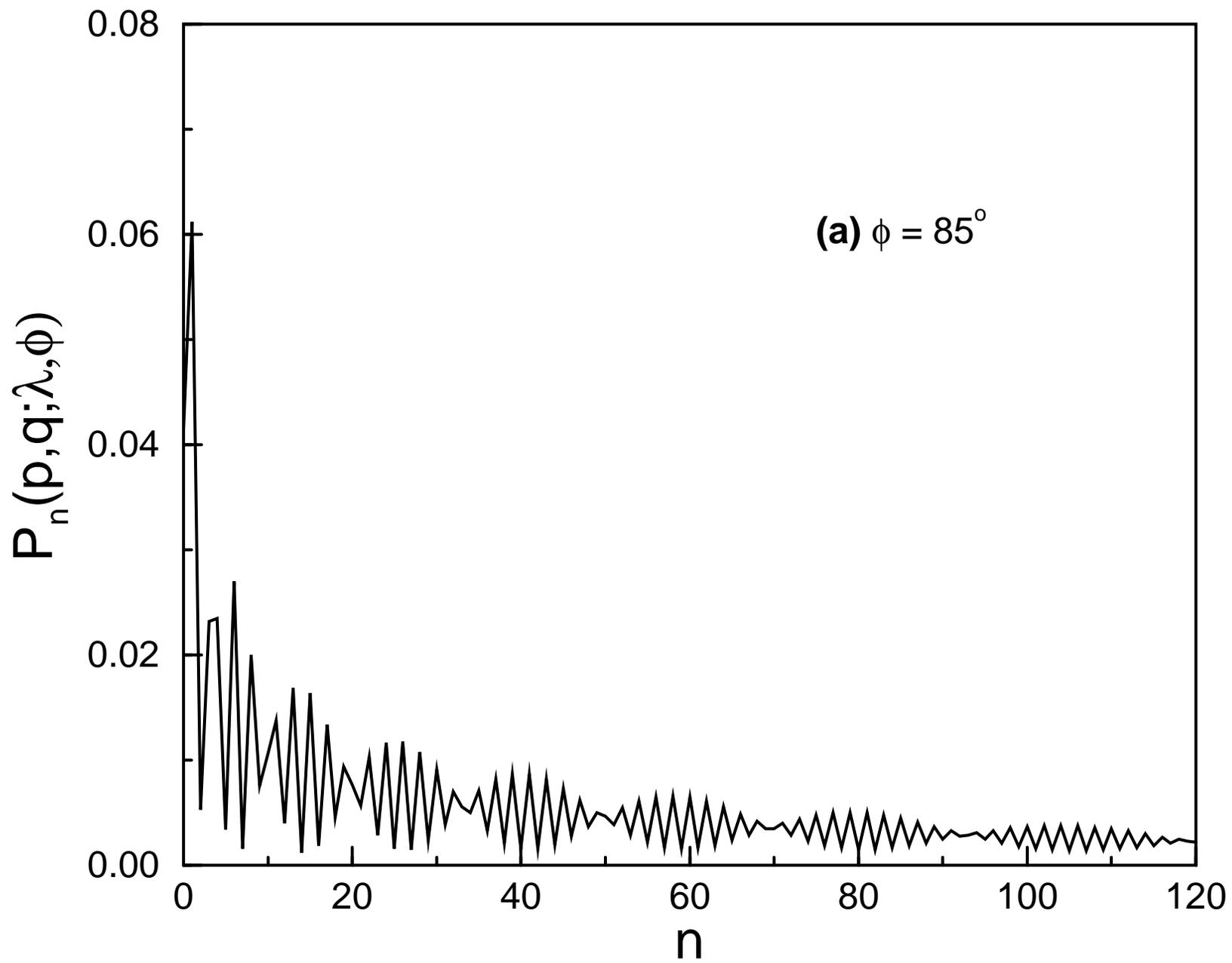
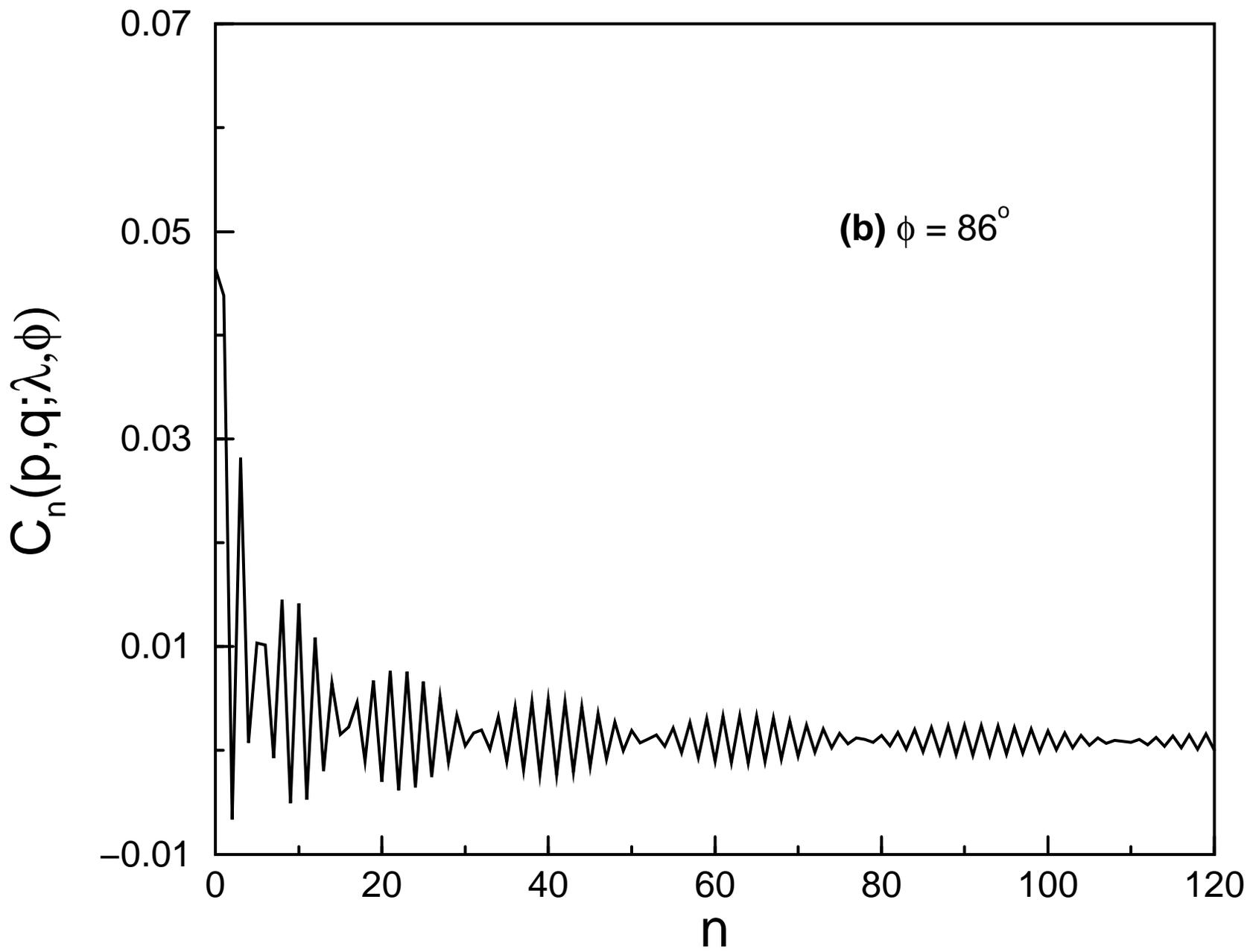
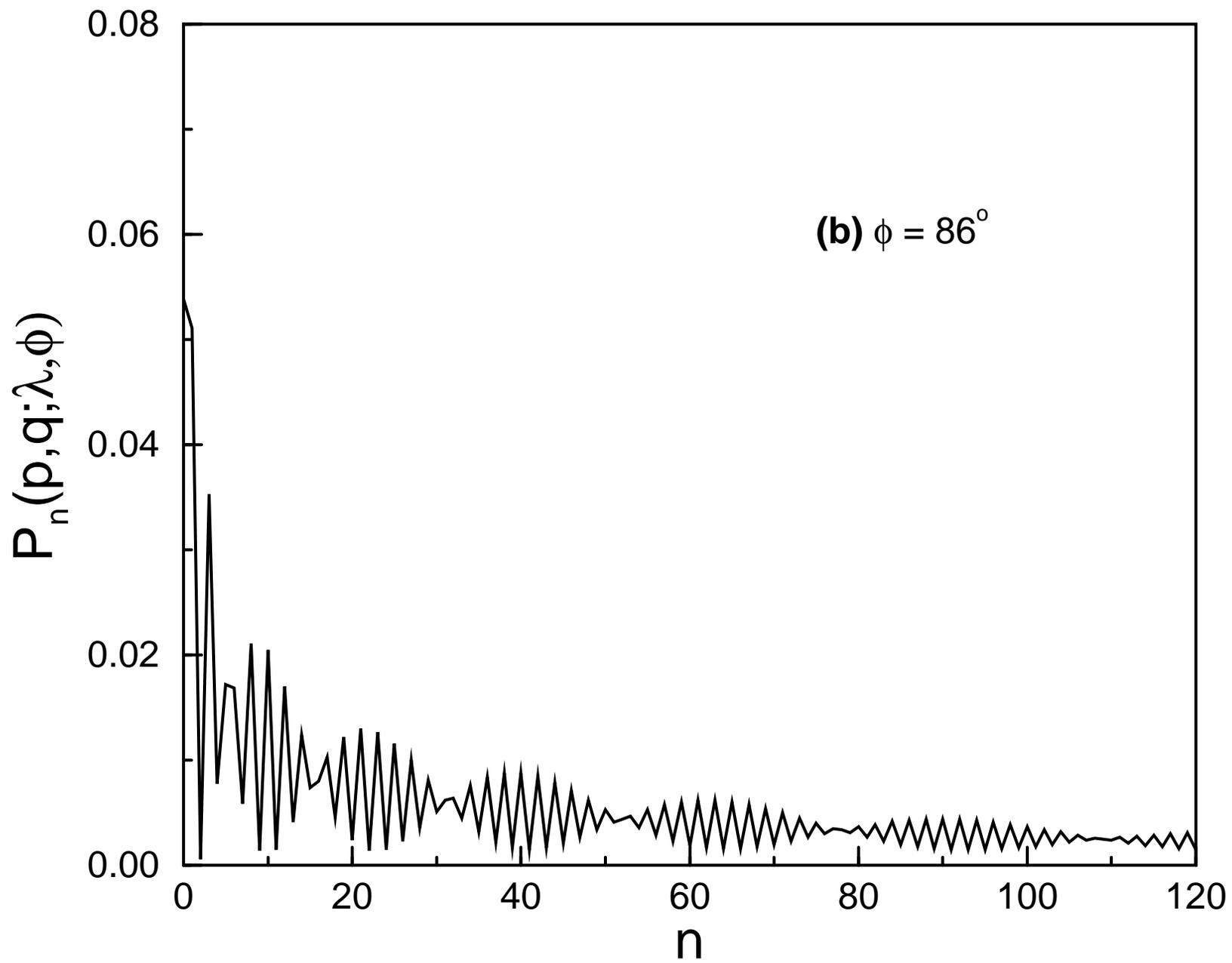
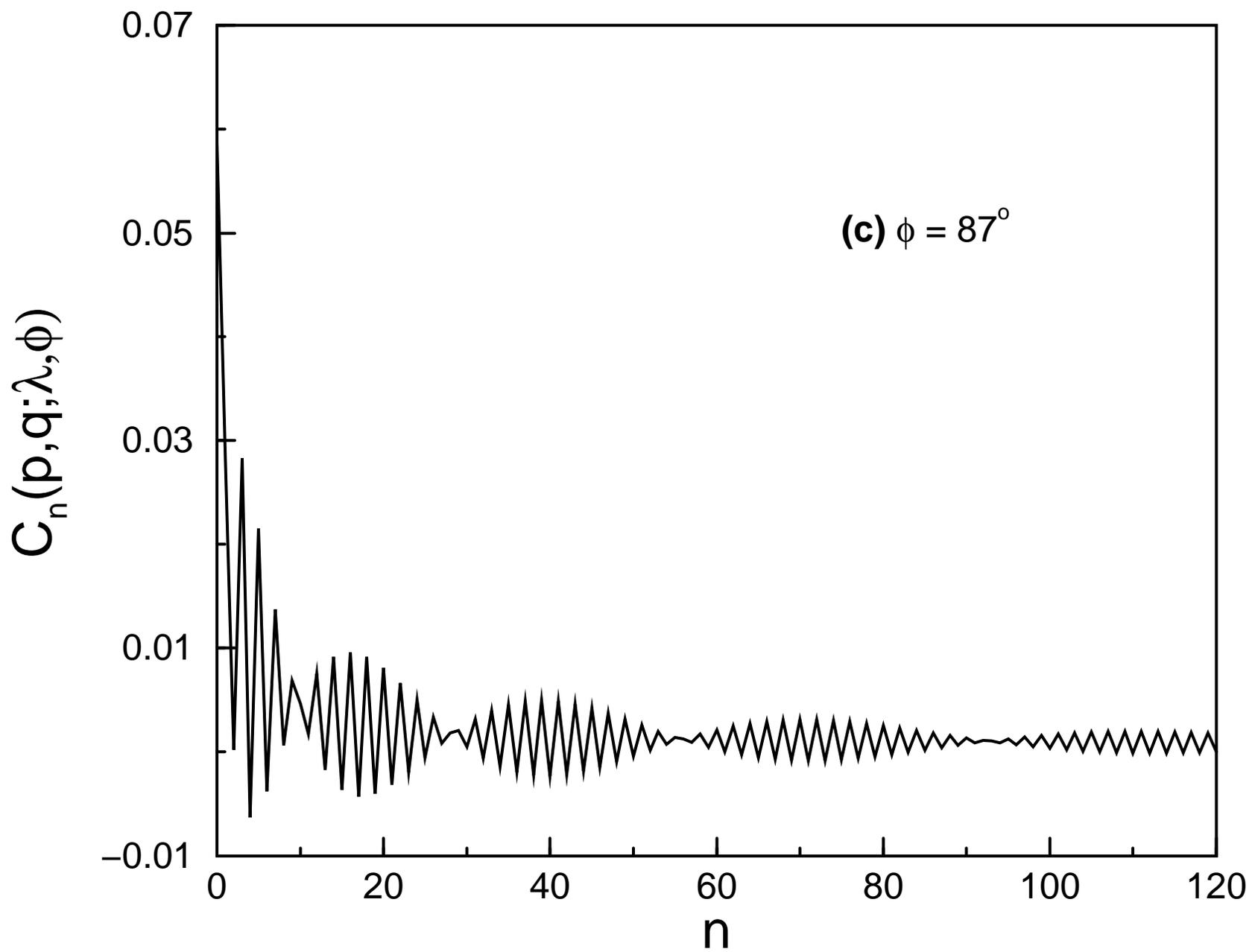
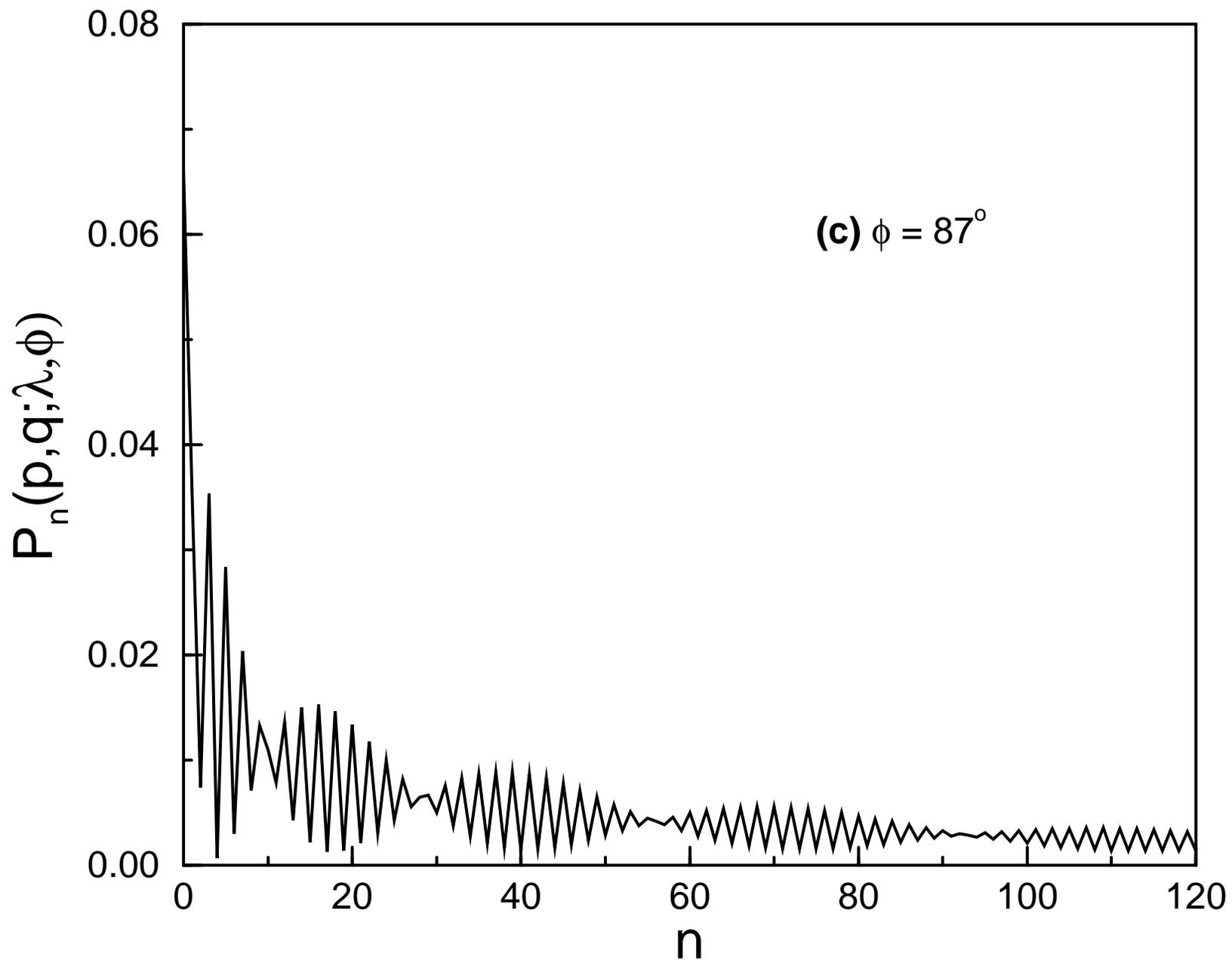
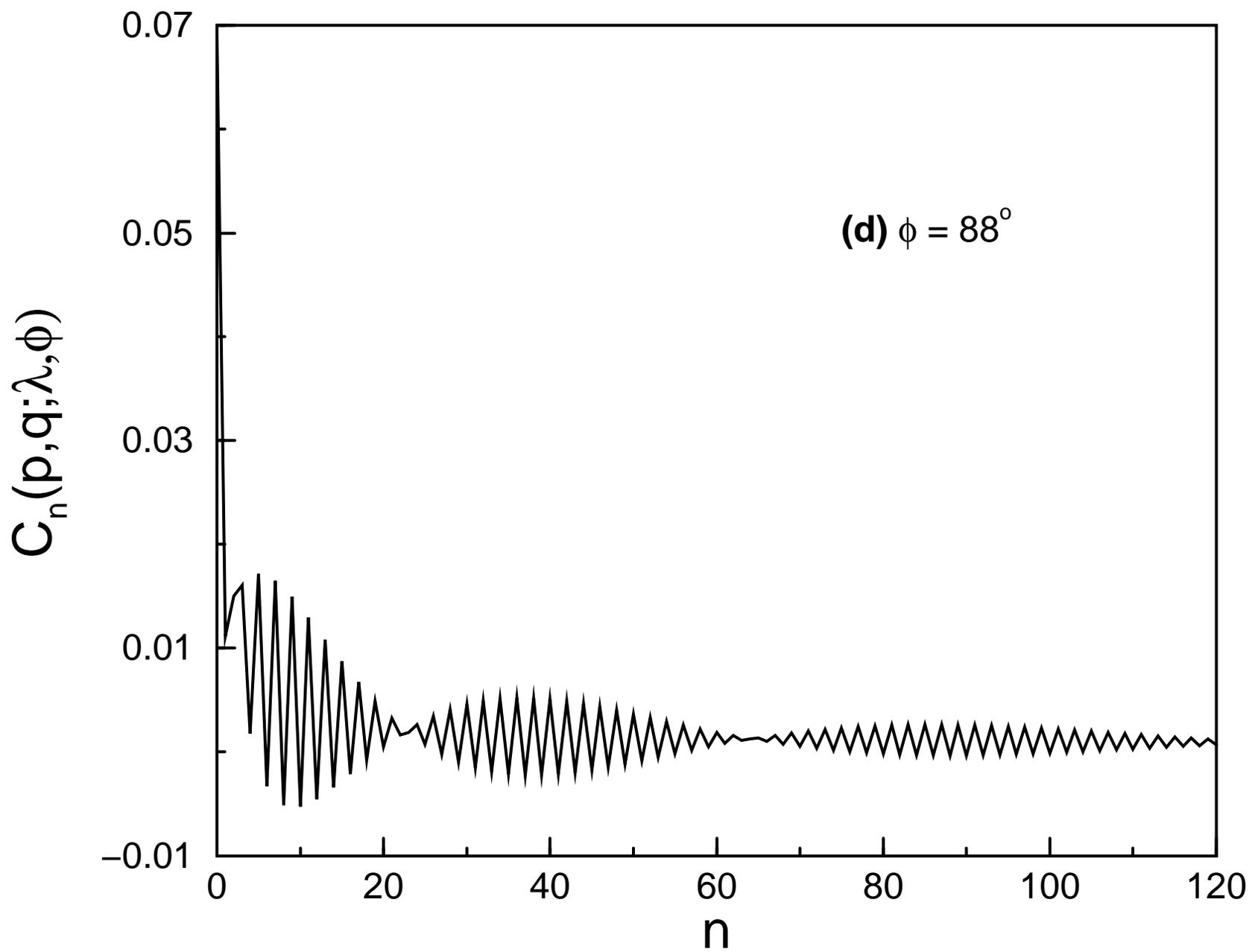
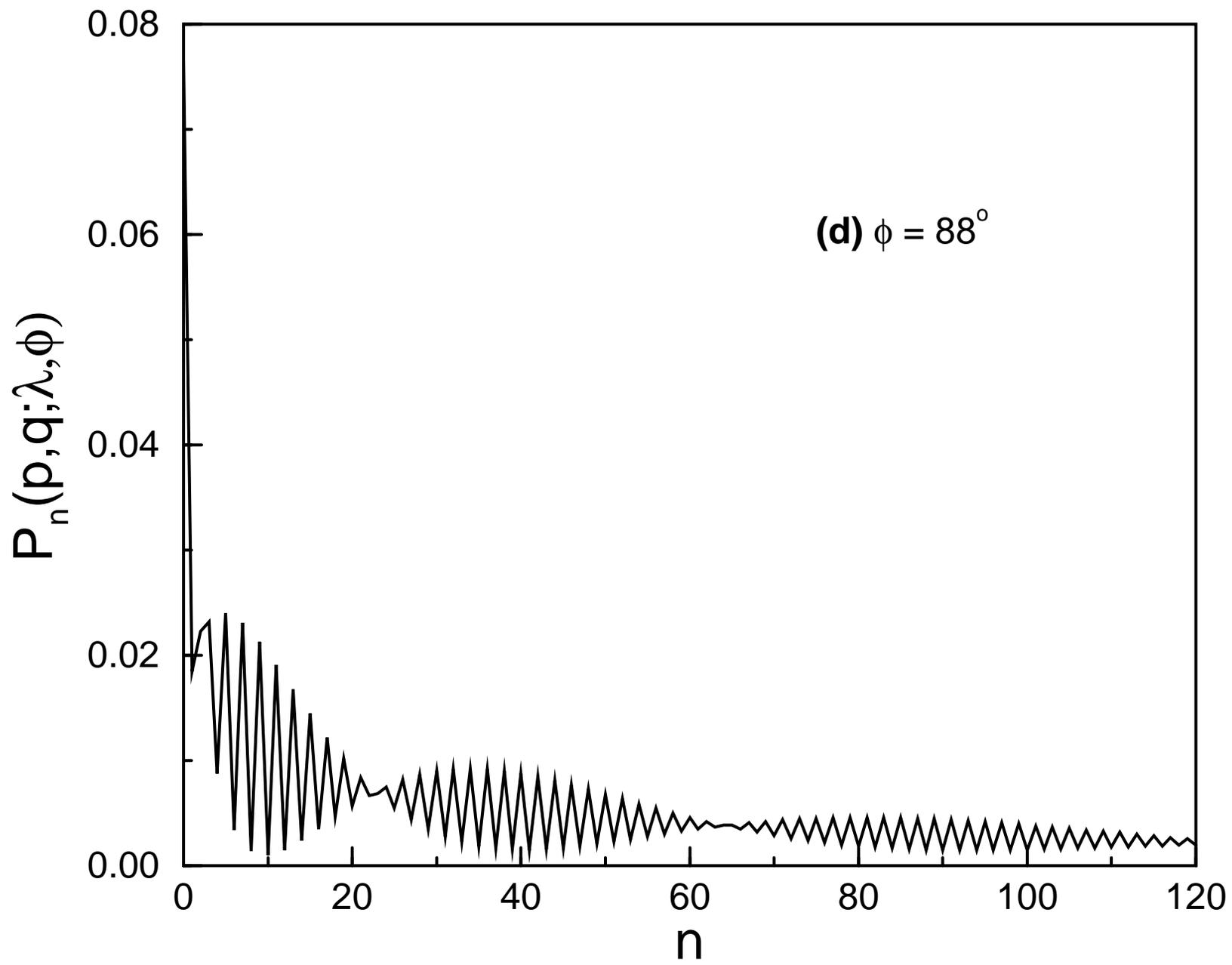
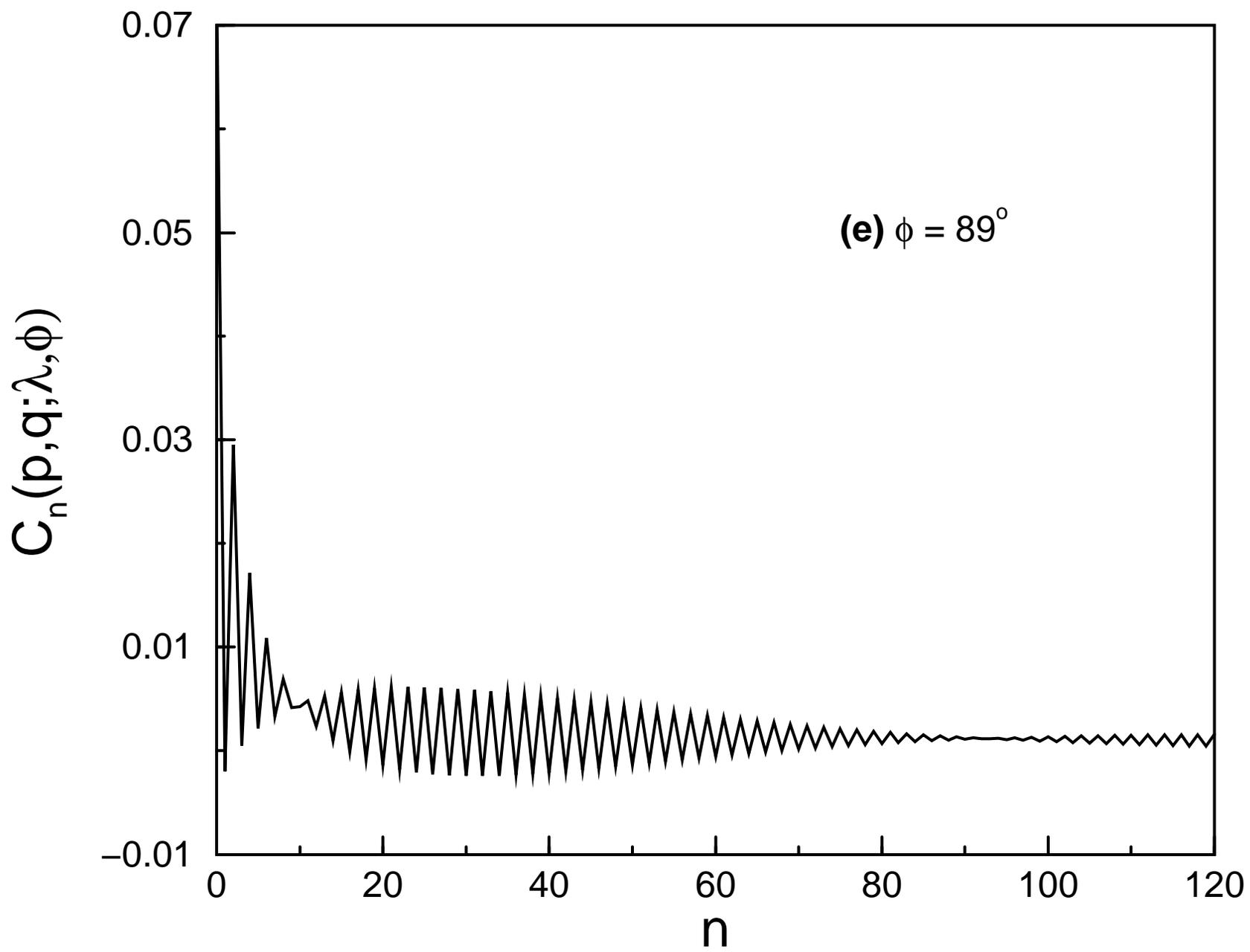
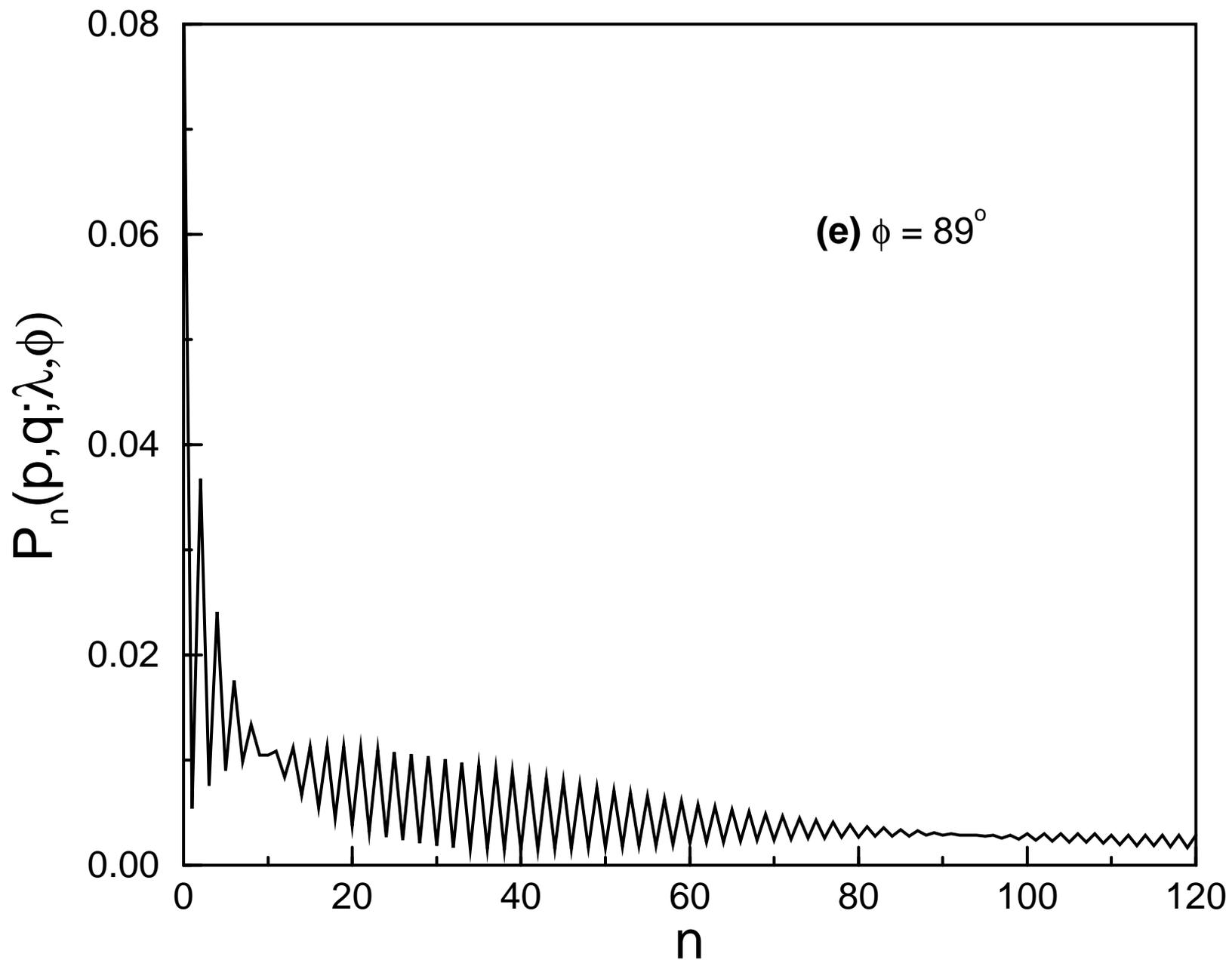
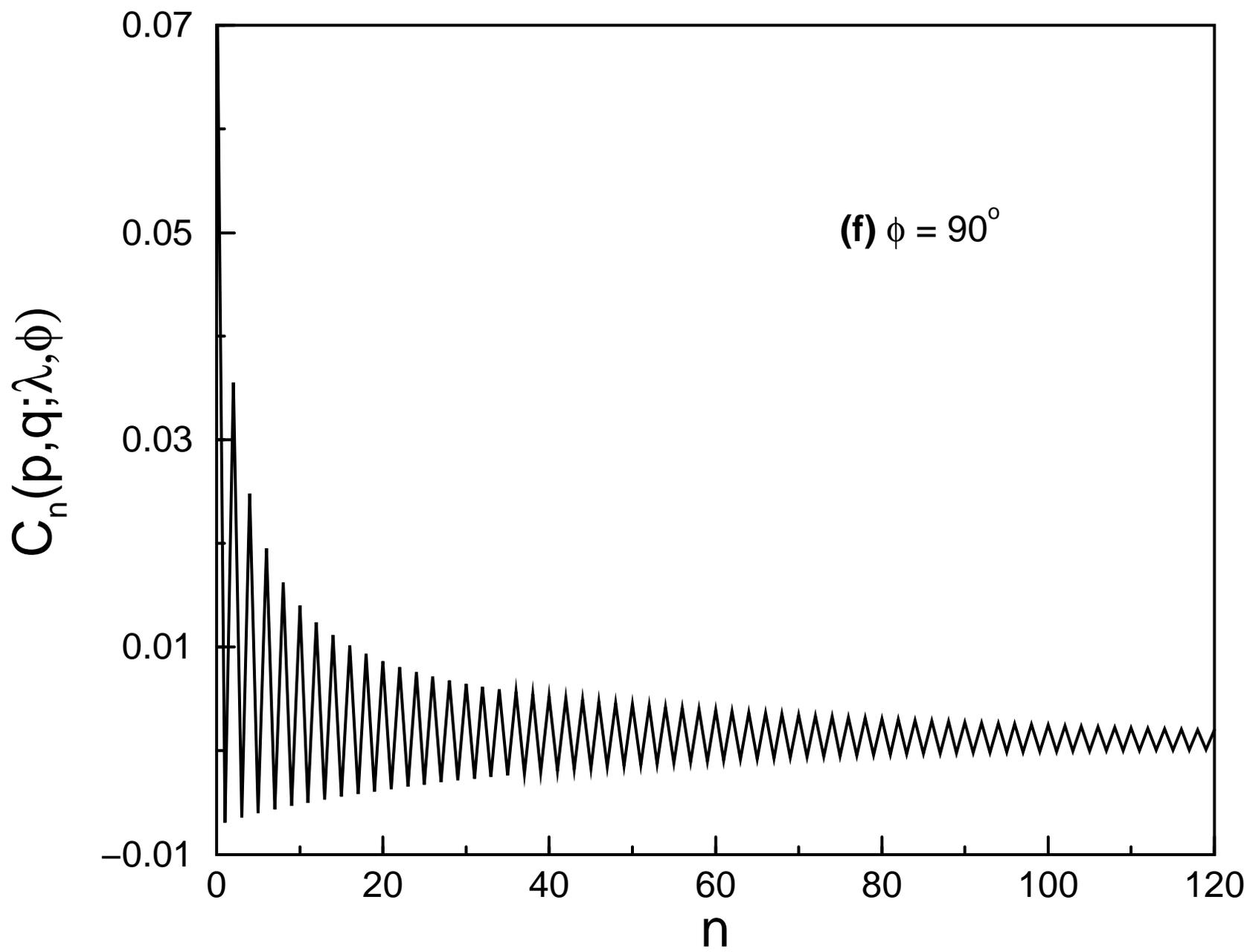
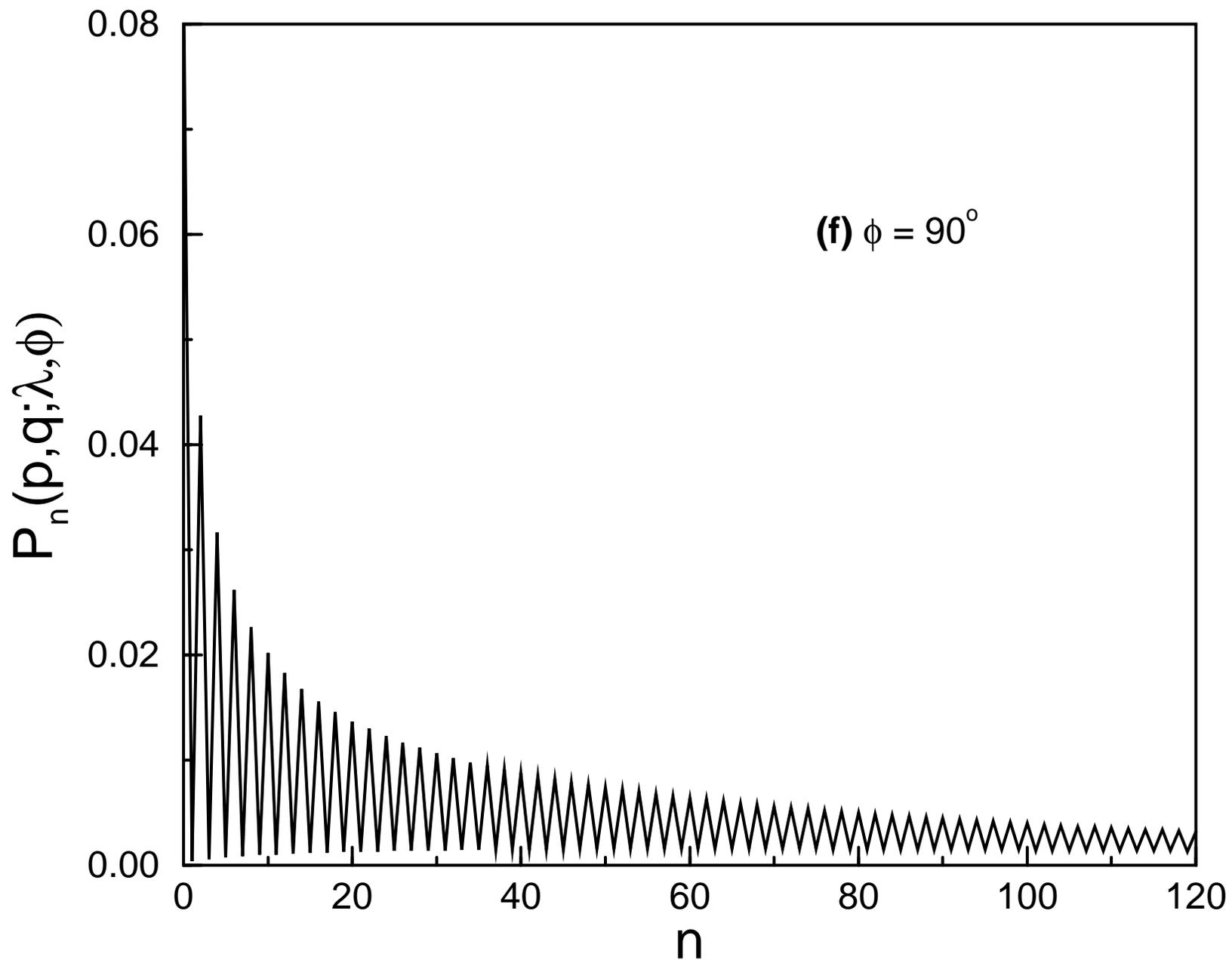
\end{document}